\documentclass[10pt,letterpaper,twocolumn]{article} %% two column, final layout

\usepackage{ol2}
\usepackage[draft]{hyperref}
\usepackage{graphicx}
\usepackage{amsmath}
\usepackage{textcomp}
\usepackage{mathrsfs}
\usepackage{amsfonts}

\begin{document}

%%%new commands
\newcommand{\beq}{\begin{equation}}
\newcommand{\eeq}{\end{equation}}
\newcommand{\barr}{\begin{eqnarray}}
\newcommand{\earr}{\end{eqnarray}}
\newcommand{\Etilde}[2]{A_{#1}(\mathbf{#2})}
\newcommand{\EtildeR}[2]{A_{#1}(\mathbf{#2})}
\newcommand{\EtildeC}[2]{A^*_{#1}(\mathbf{#2})}
\newcommand{\Atilde}[1]{\tilde{#1}(\mathbf{k},t)}
\newcommand{\campo}[1]{\mathbf{#1}(\mathbf{r},t)}
\newcommand{\vett}[1]{\mathbf{#1}}
\newcommand{\uvett}[1]{\hat{\vett{#1}}}
\newcommand{\bra}[1]{\langle #1|}
\newcommand{\ket}[1]{|#1\rangle}
\newcommand{\braket}[2]{\langle #1|#2\rangle}
\newcommand{\baseE}[2]{\uvett{e}_{#1}(\vett{#2})}
\newcommand{\baseEL}[2]{\uvett{e}_{#1}^L(\vett{#2})}
\newcommand{\baseER}[2]{\uvett{e}_{#1}^R(\vett{#2})}
\newcommand{\baseEI}[2]{\uvett{e}_{#1}^I(\vett{#2})}
\newcommand{\baseEC}[2]{\uvett{e}_{#1}^*(\vett{#2})}

\newcommand{\beam}[1]{\mathbf{#1}(\mathbf{R},z)}
\newcommand{\GH}{Goos-H\"anchen }
\newcommand{\IF}{Imbert-Fedorov }

\twocolumn[
\title{Goos-H\"anchen and Imbert-Fedorov shifts for paraxial X-Waves}

\author{Marco Ornigotti$^{1,*}$, Andrea Aiello$^{2,3}$ and Claudio Conti$^{4,5}$}	

\address{$^1$Institute of Applied Physics, Friedrich-Schiller University, Jena, Max-Wien Platz 1, 07743 Jena, Germany\\
$^2$ Max Planck Institute for the Science of Light, G$\ddot{u}$nther-Scharowsky-Strasse 1/Bau24, 91058 Erlangen,Germany\\
$^3$Institute for Optics, Information and Photonics, University of Erlangen-Nuernberg, Staudtstrasse 7/B2, 91058 Erlangen, Germany\\
$^4$ Institute for Complex Systems (ISC-CNR), Via dei Taurini 19, 00185, Rome, Italy\\
$^5$  Department of Physics, University Sapienza, Piazzale Aldo Moro 5, 00185, Rome, Italy\\
$^*$Corresponding author: marco.ornigotti@uni-jena.de}
\begin{abstract}
We present a theoretical analysis for the \GH and \IF shifts experienced by an X-wave upon reflection from a dielectric interface. We show that the temporal chirp, as well as the bandwidth of the X-wave directly affect the spatial shifts in a way that can be experimentally observed, while the angular shifts do not depend on the spectral features of the X-Wave. A dependence of the spatial shifts on the spatial structure of the X-wave is also discussed.
\end{abstract}


]

\maketitle

Nonspecular reflection phenomena, i.e., deviations from the geometrical law of reflection for light beams resulting in an effective beam shift at the interface, represented a growing field of research that produced a lot of literature in the last decade. Among all these effects, the most famous ones are the \GH \cite{ref2,ref2a, ref2b} and \IF \cite{ref2e, ref2f, ref2h,res1,res2,res3,res4,res5,res6,res7,res8} shifts, the former occurring in the plane of incidence of light, while the latter occurs in the plane orthogonal to the plane of incidence.  These phenomena have been extensively studied in the past for a vast category of beam configurations \cite{ref7a,ref8a,ref9,ref9a} and interfaces \cite{ref10a, ref10,ref11,ref12}. A comprehensive review on beam shift phenomena can be found in Ref. \cite{res9}. Although these effects are typically very small, their magnitude can be significatively enhanced by means of weak measurement-like techniques \cite{weak1,weak2,weak3,res10,res11,res12}. Very recently, the formalism typically used to calculate these beam shifts has been extended to the non-monochromatic case \cite{nostro}, giving the possibility to study how the spectral features of ultrashort pulses can influence these shifts. 

Among the vast class of non-monochromatic solutions of Maxwell's equation, on the other hand, electromagnetic X-waves are particular solutions propagating at a given superluminal velocity $v=c/\cos\beta_0$ without exhibiting diffraction or dispersion, even in the presence of a frequency dependent refractive index \cite{refX1}. This kind of solutions has been investigated in a variety of different frameworks and within an interdisciplinary perspective, encompassing the originally considered acoustic X-waves \cite{refX2}, to nonlinear optics \cite{refX3}, up to recent proposal in Bose-Einstein condensates \cite{refX4}. These light-bullets have been also recently investigated within the framework of second quantized Maxwell equations, shading new light on non-monochromatic field quantization \cite{refQuant}. In a more general settlement, X-waves are considered a part of a large family of exact solutions of the time dependent wave equation, which are commonly referred as ``localized waves" \cite{refX1}. In certain respects, X-waves are a paradigmatic example of the way a non-trivial space-time coupling may lead to an interesting propagation-invariant (solitonic-like) behavior, even in vacuum. X-waves are indeed non-monochromatic superpositions of Bessel beams with a fixed conical angle $\beta_0$. The fact that the composing Bessel beams do not exhibit diffraction and all have the same phase velocity $v$ in the propagation direction originates the progressive undistorted evolution of the X-waves. 

In this Letter, we report on the theoretical investigation of the \GH and \IF shifts for X-waves. The interaction of X-waves with dielectric interfaces have been recently studied, with particular emphasis on the superluminal tunneling \cite{superluminal}, and the occurrence of beam shifts for the case of normal incidence only \cite{ref8}.  Here we derive closed form expressions showing that the temoral chirp, the bandwidth and the angular aperture of the X-waves directly affect the \GH and \IF shifts, in a way that can be directly experimentally investigated. Furthermore, we show that the shifts are also affected by the order of the X-waves, i.e., by their specific spatial structure, a circumstance that allows to foresee the possibility of using the reflection at an interface as a prism-like effect decomposing the propagation invariant beams in the subset given by the so-called fundamental X-waves. This can be eventually used for encoding quantum information, as it happens for the polarization or for the angular momentum of Laguerre-Gauss beams \cite{OAM}.

\begin{figure}[!t]
\begin{center}
\includegraphics[width=0.5\textwidth]{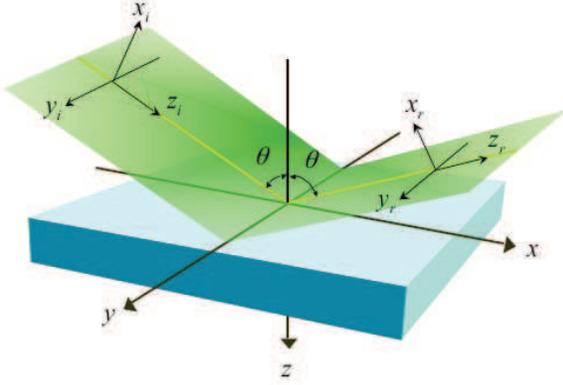}
\caption{Geometry of the problem. $\{\uvett{x}_i,\uvett{y}_i,\uvett{z}_i\}$ is the reference frame attached to the incident wave packet, $\{\uvett{x}_r,\uvett{y}_r,\uvett{z}_r\}$ is the reference frame attached to the reflected wave packet and $\{\uvett{x},\uvett{y},\uvett{z}\}$ is the laboratory frame. $\theta$ is the angle of incidence.}
\label{figura1}
\end{center}
\end{figure}
 Let us start our analysis by considering a paraxial, non-monochromatic electric field that impinges onto a dielectric surface. According to Fig. \ref{figura1}, we can define three cartesian reference frames: one attached to the incident beam ($\{\uvett{x}_i,\uvett{y}_i,\uvett{z}_i\}$) one attached to the reflected beam ($\{\uvett{x}_r,\uvett{y}_r,\uvett{z}_r\}$) and the laboratory reference frame $\{\uvett{x},\uvett{y},\uvett{z}\}$ attached to the dielectric interface. This last reference frame is chosen in such a way that the $z$-axis is normal to the interface and it is directed towards the dielectric surface and $\uvett{z}_i=\sin\theta\uvett{x}+\cos\theta\uvett{z}$, being $\theta$ the angle of incidence. The electric field in the incident reference frame can be then written by using the Fourier decomposition in plane waves as follows \cite{mandel}:
\beq\label{uno}
\vett{E}^I(\vett{r},t)=\sum_{\lambda=1}^2\int \frac{d\omega d^2K}{(2\pi)^{3/2}}\mathcal{A}_{\lambda}(\vett{K},\omega)e^{i(\vett{k_i}\cdot\vett{r}-\omega t)}+\text{c.c.},
\eeq
where $\vett{k}_i=k_0(U\uvett{x}_i+V\uvett{y}_i+W\uvett{z}_i)=k_x\uvett{x}+k_y\uvett{y}+k_z\uvett{z}$ (being $k_0=\omega/c$ the vacuum wave vector), $d^2K=dUdV$, $\text{c.c.}$ stands for complex conjugate, $\lambda$ runs over the polarization degrees of freedom of the field (with $\lambda=1$ corresponding to $p$ polarization and $\lambda=2$ to $s$ polarization respectively) and
\beq\label{amplitude}
\mathcal{A}_{\lambda}(\vett{K},\omega)=k_0^2\uvett{e}_{\lambda}(U,V,\theta)\mathcal{E}_{\lambda}(U,V,\omega).
\eeq
The polarization vectors $\uvett{e}_{\lambda}(\vett{k})$ are defined as $\uvett{e}_1(\vett{k})=(\uvett{e}_2\times\vett{k})/|\uvett{e}_2\times\vett{k}|$ and $\uvett{e}_2(\vett{k})=(\uvett{z}\times\vett{k})/|\uvett{z}\times\vett{k}|$, being $\uvett{z}$ a unit vector directed along the laboratory axis $z$. Note that for the paraxial case one has $k_z=\sqrt{k_0^2-k_x^2-k_y^2}$ and therefore $W=\sqrt{1-U^2-V^2}$. 

The quantities $\mathcal{E}_{\lambda}(U,V,\omega)$ in Eq. \eqref{amplitude} determine the shape and the polarization of the field and can be written as $\mathcal{E}_{\lambda}(U,V,\omega)=\alpha_{\lambda}(U,V,\theta)A(U,V,\omega)$, where $A(U,V,\omega)$ is the scalar spectral amplitude of the field and  $\alpha_{\lambda}(U,V,\theta)=\uvett{e}_{\lambda}(U,V,\theta)\cdot\uvett{f}$ is the vector spectral amplitude, that sets the polarization of the field in the local incident frame \cite{brewster, res7}. Here we assume that the field has passed through a polarized, whose orientation is represented by the complex-valued unit vector $\uvett{f}=f_p\uvett{x}_i+f_s\uvett{y}_i$, with $|f_p|^2+|f_s|^2=1$. 

Upon reflection each plane wave $\uvett{e}_{\lambda}(\vett{k})\exp{[i\vett{k}\cdot\vett{r}]}$ transforms according to $\uvett{e}_{\lambda}(\vett{k})\rightarrow r_{\lambda}(\vett{k})\uvett{e}_{\lambda}(\tilde{\uvett{k}})\exp{[i\tilde{\vett{k}}\cdot\vett{r}]}$, where $r_{\lambda}(\vett{k})$ are the Fresnel coefficients for $p$ and $s$ polarization respectively \cite{born} and $\tilde{\vett{k}}=\vett{k}-2\uvett{z}(\uvett{z}\cdot\vett{k})$ is determined by the law of specular reflection \cite{specular}.  The electric field in the reflected frame can be then written in a similar form as given by Eq. \eqref{uno} by substituting $\uvett{e}_{\lambda}(\vett{k})\rightarrow\uvett{e}_{\lambda}(\tilde{\vett{k}})\equiv\uvett{e}_{\lambda}(-U,V,\pi-\theta)$ and $\vett{k}_i\rightarrow\vett{k}_r=-U\uvett{x}+V\uvett{y}+W\uvett{z}$. The explicit expressions of $\uvett{e}_{\lambda}(U,V,\theta)$, $\uvett{e}_{\lambda}(\tilde{\vett{k}})$, $r_{\lambda}(U,V)$ and $\alpha_{\lambda}(U,V,\theta)$ for a paraxial beam are given in Ref. \cite{brewster}. 

According to the formalism developed in Refs. \cite{andreaNJP,nostro}, we calculate the first order moment of the intensity distribution of the reflected field (within the paraxial approximation) as
\beq\label{res1eq}
\langle\vett{X}\rangle=\langle x_r\rangle\uvett{x}_r+\langle y_r\rangle\uvett{y}_r,
\eeq
where
\beq\label{centroid}
\langle\xi\rangle=\frac{\int dt\;d^2R\;\xi\;\mathcal{I}(x_r,y_r,z_r,t)}{\int dt\;d^2R\;\mathcal{I}(x_r,y_r,z_r,t)},
\eeq
where $\xi=\{x_r,y_r\}$, $d^2R=dx_rdy_r$ and (up to an inessential multiplicative constant) $\mathcal{I}(x_r,y_r,z_r,t)=\vett{E}^R(x_r,y_r,z_r,t)\cdot\vett{E}^R(x_r,y_r,z_r,t)$, where $\vett{E}^R(x_r,y_r,z_r,t)$ is the electric field in the reflected frame and is given by Eq. \eqref{uno} with the aforementioned substitutions. 

Using Eq. \eqref{res1eq} it is possible to express both the spatial ($\Delta$) and angular ($\Theta$) \GH and \IF shifts as 
\begin{subequations}\label{shifts1}
\begin{align}
\Delta_{GH}&=\langle x_r\rangle|_{z=0},\\
\Theta_{GH}&=\frac{\partial\langle x_r\rangle}{\partial z},\\
\Delta_{IF}&=\langle y_r\rangle|_{z=0},\\
\Theta_{IF}&=\frac{\partial\langle y_r\rangle}{\partial z}.
\end{align}
\end{subequations}

In order to calculate these expressions, however, one needs to know the form of the scalar spectral amplitude $A(U,V,\omega)$ that appears in the definition of $\mathcal{E}_{\lambda}(U,V,\omega)$. For the case of X-waves, in general, the spectral dependence can be factorized from the spatial dependence, leading to $A(U,V,\omega)=\gamma(\omega)\mathcal{S}(U,V)$. Since X-waves are superpositions of Bessel beams, the term $\mathcal{A}(U,V)$ is essentially the angular spectrum of a monochromatic Bessel beam.  By using spherical coordinates for the $k$-vector components, namely $\{U,V,W\}=\{\sin\beta\cos\phi,\sin\beta\sin\phi,\cos\beta\}$ it is possible to write such an angular spectrum as follows  \cite{ref12, angularX}:
\beq\label{angularS}
\mathcal{S}(U,V)\equiv\mathcal{A}(\beta,\phi)=e^{i\Big(n\phi-n\frac{\pi}{2}\Big)}\frac{\delta(\beta-\beta_0)}{|\sin\beta_0|\cos\beta_0},
\eeq
where $n$ is the order of the Bessel function that appears in the integral definition of the X-wave and $\beta_0$ is the cone aperture of the Bessel beam and the half angle between the orientation of the X-branches \cite{angularX}. 
Using then the ansatz $A(U,V,\omega)=\gamma(\omega)\mathcal{S}(U,V)$ and the form of the angular spectrum $\mathcal{S}(U,V)$ as given by Eq. \eqref{angularS}, the beam shifts for an X-wave can be written as follows:
\begin{subequations}\label{shiftsX}
\begin{align}
\Delta_{GH}&=\Big(\frac{\Omega_N}{\Omega_D}\Big)\Delta_{GH}^{(B)},\\
\Theta_{GH}&=\Theta_{GH}^{(B)},\\
\Delta_{IF}&=\Big(\frac{\Omega_N}{\Omega_D}\Big)\Delta_{IF}^{(B)},\\
\Theta_{IF}&=\Theta_{IF}^{(B)},
\end{align}
\end{subequations}
where
\begin{subequations}\label{omegas}
\begin{align}
\Omega_N&=\frac{1}{c}\int\omega|\gamma(\omega)|^2d\omega,\\
\Omega_D&=\frac{1}{c^2}\int\omega^2|\gamma(\omega)|^2d\omega,
\end{align}
\end{subequations}
and $\Delta_{GH,IF}^{(B)}$ are the \emph{adimensional} spatial shifts of a monochromatic Bessel beam defined as follows \cite{ref12}:
\begin{subequations}\label{SpatBess}
\begin{align}
\Delta_{GH}^{(B)}&=\left(w_p\frac{\partial\phi_p}{\partial\theta}+w_s\frac{\partial\phi_s}{\partial\theta}\right)\nonumber\\
&-n\frac{w_pa_s^2-w_sa_p^2}{a_pa_s}\cos\eta\cot\theta,\\
\Delta_{IF}^{(B)}&=-\frac{w_pa_s^2-w_sa_p^2}{a_pa_s}\sin\eta\cot\theta\nonumber\\
&-2\sqrt{w_pw_s}\sin(\eta-\phi_p-\phi_s)\cot\theta\nonumber\\
&-n\left(w_p\frac{\partial\ln R_p}{\partial\theta}+w_s\frac{\partial\ln R_s}{\partial\theta}\right).
\end{align}
\end{subequations}
The quantities $\Theta_{GH,IF}^{(B)}$ appearing in Eqs. \eqref{shiftsX} are the angular \GH and \IF shifts of a monochromatic Bessel beam, whose explicit expression is, according to Ref. \cite{ref12}, the following:
\begin{subequations}
\begin{align}
\Theta_{GH}^{(B)}&=-\sin^2\beta_0\left(w_p\frac{\partial\ln R_p}{\partial\theta}+w_s\frac{\partial\ln R_s}{\partial\theta}\right),\\
\Theta_{IF}^{(B)}&=\sin^2\beta_0\frac{w_pa_s^2-w_sa_p^2}{a_pa_s}\cos\eta\cot\theta.
\end{align}
\end{subequations}
In the previous definitions, $w_{\lambda}=(a_{\lambda}^2R_{\lambda}^2)/(a_p^2R_p^2+a_s^2R_s^2)$ is the fractional energy contained in each polarization and $a_{p,s}$ have been chosen in such a way that $f_p=a_p$ and $f_s=a_s\exp{(i\eta)}$. We moreover note that the quantity $\Omega_N/\Omega_D$ has, according to Eqs. \eqref{omegas}, the dimension of a length. This is consistent with the fact that  the spatial \GH and \IF shifts for Bessel beams in Eqs. \eqref{SpatBess} are given as dimensionless quantities.

As can be seen by direct comparison with the results obtained for a monochromatic Bessel beam \cite{ref12}, the non-monochromatic nature of X-waves reflects itself in an extra multiplicative factor in front of the spatial shifts. Thus, depending on the particular form of the X-wave considered, while the angular shifts will not be influenced (and it will be proportional to  $\sin^2\beta_0$, i.e., to the superluminal character of the X-wave), different X-waves will experience different spatial shifts. 

 In order to understand this result, let us consider, for example,  the \GH shift, given, according to Eq. \eqref{res1eq},  by $\langle x_r\rangle$. By partial integration the numerator of Eq. \eqref{centroid} can be tranformed from  $\langle x_r\rangle$ to  $\langle (ik_0)^{-1}\partial/\partial U\rangle$. When this derivative acts on the reflected electric field, all terms that do not involve the derivative of  the exponential term 
  $\exp{(ik_0Wz)}$ are multiplied by a prefactor $1/k_0$. According to Eq. \eqref{amplitude}, therefore, the contribution of the X-wave spectrum to the \GH shifts will be an extra multiplicative factor, namely $\Omega_N$. On the other hand, when calculating the derivative of the exponential term $\exp{(ik_0Wz)}$ with respect to $U$, an extra $k_0$-factor is brought down by the derivative itself. This cancels exactly the $(ik_0)^{-1}$ term in  $\langle (ik_0)^{-1}\partial/\partial U\rangle$. Therefore, the correspondent multiplicative term (that accounts for X-wave spectrum) associated to the numerator of the $z$-dependent part of the \GH shift is given by $\Omega_D$. Upon normalization, since the denominator of Eq. \eqref{centroid} contains $\Omega_D$, this term will simplify the correspondent term in front of the $z$-dependent shift, thus making it unaffected by the non-monochromatic nature of the field here considered. The same explanation is also valid for the \IF shift.
 
As an example, we now apply this result to two different classes of X-waves, namely the  fundamental X-waves, that posses an exponentially decaying spectrum, and the Bessel-X pulses, whose spectrum is Gaussian. For the former case, the spectral function $\gamma(\omega)$ is given by
\beq\label{gammaX}
\gamma(\omega)=(a_0\omega)^me^{-a_0\omega}e^{-im\frac{\pi}{2}}H(\omega),
\eeq
where $a_0$ (that has the dimensions of a time) is the pulse duration, $m$ is the order of the X-wave and $H(\omega)$ is the Heaviside step function. The multiplicative factor in the spatial shifts in Eqs. \eqref{shiftsX} is then given by
\beq
\frac{\Omega_N}{\Omega_D}=\frac{a_0c}{1+m}.
\eeq
%This result is quite interesting, since the spatial \GH and \IF can be reduced by increasing the order $m$ of the X-wave, while they experience an amplification proportional to $a_0$. If we assume to use the subset of fundamental X-waves as a basis for representing propagating invariant beams, then the reflection onto an interface acts like a prism for the different components, shifting different fundamental X-waves (each with a different $m$) by a different amount, proportional to their spectrum and order. This is the second result of our paper.
As a more realistic example, we consider a Bessel-X pulse, whose bounded Gaussian spectrum makes it experimentally reproducible, in contrast with the unbounded spectrum of fundamentals X-wave that gives, as a consequence, waves that carry infinite energy \cite{refX1}. For a Bessel-X pulse, therefore, the spectral amplitude $\gamma(\omega)$ is given by
\beq\label{gammaBX}
\gamma(\omega)=\frac{T_0}{\sqrt{2\pi(1+i\xi)}}e^{-q^2(\omega-\omega_0)^2},
\eeq
where $q^2=T_0^2/[2(1+i\xi)]$, $T_0$ is the pulse duration, $\xi$ is the pulse chirp parameter and $\omega_0$ is the central frequency of the Bessel-X pulse. With this choice of spectrum, the factor $\Omega_N/\Omega_D$ in Eqs. \eqref{shiftsX} assumes the following from:
\beq
\frac{\Omega_N}{\Omega_D}=\frac{2cT_0^2\omega_0}{1+\xi^2+2T_0^2\omega_0^2}.
\eeq
The spatial \GH and \IF shifts can be then controlled by tuning the pulse duration $T_0$, the pulse chirp $\xi$ and the pulse frequency $\omega_0$.

In conclusion, we have calculated the \GH and \IF shifts for a paraxial X-wave, showing how the polychromatic nature of these beams affects the shift by introducing a multiplicative factor on the spatial shifts $\Delta_{GH,IF}$ and leaving the angular shifts $\Theta_{GH,IF}$ unchanged. This extra factor is, for the case of fundamental X-waves, directly connected to the X-wave order parameter $m$ and its bandwidth $a_0^{-1}$. We then suggested to use beam shifts as an active method for analyzing the spectral content of localized fields in terms of fundamental X-waves. Last, but not least, we have also shown how the \GH and \IF shifts are affected in the case of Bessel-X pulses, showing that the pulse duration of such optical pulses can be used as a tuning parameter to control the beam shifts.

\end{document}